# Towards a Live Anonymous Question Queue To Address Student Apprehension

Lloyd Montgomery, Guy Evans, Francis Harrison, Daniela Damian
Department of Computer Science,
Software Engineering Global interAction Lab (SEGAL)
University of Victoria, Canada
{lloydrm, gwevans, francish, danielad}@uvic.ca

## ABSTRACT
In today's university climate many first and second year classes have over a hundred students. Large classrooms make some students apprehensive about asking questions. An anonymous method of submitting questions to an instructor would allow students to ask their questions without feeling apprehensive. In this paper we propose a Live Anonymous Question Queue (LAQQ), a system that facilitates anonymous question submissions in real time to mitigate student apprehension, increase student participation, and provide real-time feedback to the instructor. To study the necessary features of an LAQQ, we conducted a study of a system, namely Google Moderator, which best approached our concept of an LAQQ. We deployed Google moderator in large lectures and studied its support of a number of features that we envisioned for an LAQQ. Through our class observations, interviews with instructors, and surveys with the students, our results suggest that an LAQQ system must provide support for: notification of question submission to provide awareness for the instructor, and context for questions to allow an instructor to easily answer a question. Additionally our results suggest that an LAQQ system must be accessible and usable on multiple platforms. Finally our results suggest that in order to be successful in the classroom an LAQQ system must be fully adopted by the instructor and the classroom organizational structure must change to accommodate the use of the LAQQ.

## Categories and Subject Descriptors
K.3.1 [**Computers and Education**]: Computer Uses in Education – *collaborative learning, computer-assisted instruction, and computer-managed instruction.*

## Keywords
Live Anonymous Question Queue, Student Apprehension, Computer Supported Education

## 1. INTRODUCTION
In today's university climate many first and second year classes have over a hundred students. Large classrooms make some students apprehensive about asking questions for fear of negative evaluation from their peers [3]. Evaluation apprehension is the fear of negative evaluation from peers in a group and is a known source of group process loss [5]. Advances in computing have created opportunities to use technology to allow a new line of communication between students and their instructors.

In this paper we report on our research into a live anonymous question queue (LAQQ) system that facilitates anonymous question submissions in real time, to mitigate student apprehension, increase participation and provide real—time feedback to the instructor. To study features necessary for an LAQQ, we conducted a study of Google Moderator (GM), a tool that fit our model of an LAQQ. Our study consisted of interviewing instructors, observing Google Moderator use in classrooms, surveying the students involved, and interviewing the instructors post-study. We hypothesized that a well-accepted implementation of an LAQQ into the classroom setting would increase student participation in the classroom, increase the knowledge that the instructor has about their student's level of understanding, and decrease the apprehension level of students.

During our study we found that the defined features an LAQQ system must support was much broader, encompassing more features than we first imagined. Through analysis of the data collected from our study, we compiled a more expansive list of the features we proposed as essential to an LAQQ as well as necessary conditions for its implementation in a classroom setting.

## 2. L.A.Q.Q.
Recent advances in computer-mediated collaboration offer the opportunity for computer technologies to be an integral part of classroom instruction, in particular in (1) mitigating student apprehension, (2) increasing student participation, and (3) providing real-time feedback to the instructor. A number of computer-supported learning tools have been developed that offer varying support to student-student or student-instructor collaboration. For example, ActiveClass [7] encouraged in-class participation through supplied personal wireless devices, Backstage [6] allowed participants to collaborate with each other and the presenter via micro-blogs, Computer-Mediated Feedback System [9] allowed students to communicate with the instructor through annotations on the slides, and FSM [2] provided students with the ability to anonymously communicate with the lecturer through expressive anonymous feedback.

Below, we briefly review the efforts these systems used to mitigate apprehension, increase participation, and provide feedback for the lecturer. We then highlight the concepts that inspired our development of LAQQ.

**Mitigating Apprehension:** Apprehension was addressed through building *anonymity* into the systems, at various levels. For example, in BackStage [6] and FSM [2] students were able to anonymously ask their question, while in ActiveClass [7] the input was pseudo-anonymous because the professor was able to identify authors if needed.

**Increasing Participation:** Increasing student participation in classrooms has been proposed through implementing a *question queue* that was easily *accessible* to all students, such as through micro blogging to increase peer-to-peer communication [6], or on public displays in the classroom [2, 9] or on personal devices [7].

**Providing Feedback to Lecturers:** Feedback of common questions for the lecturer was implemented by *question ranking* [7]. *Live feedback* of student questions was provided to the lecturer's through a separate display [2,7,9]. Explicit *notifications* of incoming questions were posted directly on the lecturer's slide [9].

Drawing on this work, our proposed LAQQ is a system to mitigate apprehension, increase participation, and provide feedback to the instructor that supports the following features:

- **Live Feedback**: to provide real-time updates to the instructor
- **Anonymity**: to mitigate student apprehension
- **Question Queue**: to organize student input
- **Question-ranking**: to rank common questions
- **Notifications**: to keep students and instructors aware
- **Accessibility**: to allow all students to participate

Previous research did not specifically nor explicitly study the complete set of these features. The tools we reviewed were studied only for their support of a limited set of features – anonymity, accessibility, and increased participation in FSM [2] or pseudo-anonymity in ActiveClass [7], or were not tested in real classrooms [6]. In the remainder of this paper we describe our empirical study of a system that provides some support for all these features. By reflecting upon our observations of Google Moderator in the classroom, and interviews with the students and instructors, we were able to develop an enhanced, richer understanding of the necessary features for a LAQQ and its successful implementation in the classroom.

## 3. RESEARCH METHODOLOGY

Google Moderator (GM) supports the collection and display of student questions with a special presentation mode, which provides notification and live-feedback for the instructor by implementing a question queue. Users of GM can select an anonymous username and since GM is cross-platform it supports accessibility. Support for question ranking is provided by allowing all students to vote on questions in the question queue.

Our study followed a five step process: (1) we interviewed instructors, (2) we conducted an observational period in each classroom, (3) we conducted a field experiment with GM in each classroom, (4) we surveyed the students, and (5) we conducted post-study interviews with the instructors.

### 3.1 Participants
We found three instructors that were willing to try GM in their classes. Table 1 gives a breakdown of each section. Instructor *AB* taught two second year logic courses; instructor *C* taught a first year computer skills course; and instructor *DE* taught two sections of the same introduction to programming class.

Table 1. Study Sections

| Instructor | Class | Size | Identifier |
|---|---|---|---|
| AB | Critical Thinking | 212 | A |
|  | Formal Logic | 106 | B |
| C | Into. Computer Skills | 206 | C |
| DE | Intro Java | 183 | D |
|  |  | 177 | E |

### 3.2 Pre-Study Instructor Interviews
Different instructors use different teaching styles so we allowed the instructors to decide how they would use an LAQQ system in their classroom. We met with each instructor and demonstrated the main features of GM and asked them how they would like to use it in their courses. Our initial idea was to use GM during class, providing a laptop for each instructor to easily view presentation mode and answer questions as they came in. However, we were surprised to find all three instructors wanted to keep GM sessions open during and between classes. All of the instructors decided to collect questions during and after class to answer at the beginning of the next class.

### 3.3 Observational Period
We conducted an initial observational period in each class to see each instructor's teaching style so that we could make suggestions on how they might use GM. Instructors *AB* and *C* used a standard lecture format with slides, whereas instructor *DE* used a flipped classroom model [8], which allowed her to walk around during the lecture answering questions individually. Due to this teaching style instructor *DE* decided to use GM's presentation mode on a smartphone while they wandered.

We then deployed GM in each of the five sections for one week. For sections A, B, D, and E Google Moderator was deployed for two consecutive 1.5 hour classes each. For section C Google Moderator was deployed for three consecutive 1 hour classes. Total GM usage time for all the classes combined was 15 hours. Three researchers participated in each class, one for moderating the GM session, and two for observing student and instructor usage of GM. At the beginning of the first class for each section a brief presentation and introduction to GM was given for the students.

Laptops were provided with presentation mode open for each instructor, and instructor *DE* used their smart phone with presentation mode while conducting the flipped class. Instructor *C* gave time at the end of each class for students to submit questions to be answered at the beginning of the next class. All GM sessions were kept open between classes.

### 3.4 Survey
We wanted to find out what the students experience was like using, or not using, GM. We also wanted to find out if students were comfortable asking questions in class. We created an online survey for all students in each class to take, including those students who did not use GM during the study.

### 3.5 Post-Study Instructor Interviews
We conducted post study interviews with each instructor to get feedback on how they felt about using GM, and whether or not they would use it, or another LAQQ system, again. We also wanted to get feedback on what features of GM worked for them, what features did not work, and what features they wish the tool had.

## 4. ANALYSIS AND RESULTS

We asked every student in each class to complete our survey, including those students that did not use GM. Our survey received a total of 137 responses. In this section we present our results on student apprehension and anonymity, and then follow with our results on Google Moderator usage.

### 4.1 Student Apprehension and Anonymity

Table 2 shows the breakdown of responses to the question: "how do you feel about asking questions in class?"

**Table 2. Student comfort level**

| Very Comfortable | 8.0 % |
|---|---|
| Comfortable | 20.4 % |
| Indifferent | 30.0 % |
| Uncomfortable | 35.0 % |
| Very Uncomfortable | 6.6 % |

In our analysis of the results we grouped respondents who answered "very comfortable" and "comfortable" together, and we grouped respondents who answered "very uncomfortable" and "uncomfortable" together. The results show that 41% of our survey respondents are uncomfortable asking questions in class.

We asked the respondents in the "uncomfortable to ask questions" category if they would be more comfortable asking questions anonymously. 93% said that they would, and 7% said that they would not.

We asked the uncomfortable respondents "why are you uncomfortable asking questions in class?" and allowed them to select all that apply. We found that the number one reason students are uncomfortable asking questions in class is due to them being scared of asking a dumb question, as shown in table 3.

**Table 3. Reasons for being uncomfortable asking questions**

| English is not my first language | 3.0 % |
|---|---|
| I have trouble phrasing my questions | 12.1 % |
| I have a fear of public speaking | 22.0 % |
| I am scared of asking a dumb question | 26.5 % |
| I am worried my classmates will judge me | 22.0 % |
| I am worried my instructor will judge me | 12.9 % |
| Other | 1.5 % |

### 4.2 Google Moderator Usage

To investigate the GM usage, we analyzed the type and number of questions asked both in class and through GM. We also analyzed the survey questions that related to GM usage. Finally we investigated our own observational data.

Figure 2 shows a breakdown of all of the questions asked during the study period for each class. Questions asked *out loud* were those questions students asked the lecturer in front of the entire class, *individual* questions occurred in the flipped classes *D* and *E* and were questions that students asked the teacher on an individual basis. *GM Relevant asked* are those questions that were asked on Google Moderator that were relevant to course material, and *GM answered* were questions answered that were sourced from GM.

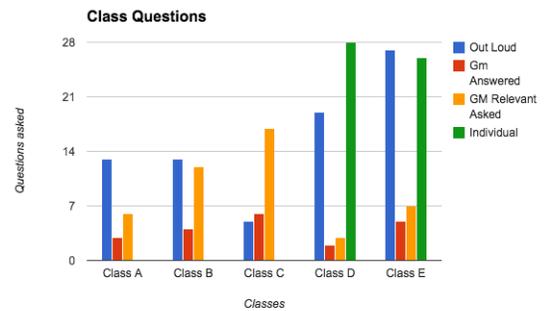

Figure 2 – Questions Asked & Submitted

#### 4.2.1 Survey Results

We asked all of our respondents if they used GM, and if so what their level of involvement was. We found that 57.2% of our respondents used GM with some level of involvement. Table 4 gives a breakdown of that involvement.

**Table 4. Highest Level of Google Moderator Involvement**

| Yes, I asked a question | 11.0 % |
|---|---|
| Yes, I voted on a question | 22.6 % |
| Yes, I viewed the questions | 23.6 % |
| No I did not use it | 43.8 % |

We then asked those respondents who used GM, but did not ask a question, why they did not ask a question (table 5).

**Table 5. Reasons for not submitting a question to GM**

| I prefer to ask my questions out loud | 9.1 % |
|---|---|
| I did not have a question | 57.1 % |
| I was preoccupied with the lecture | 26.0 % |
| I did not understand how to ask a question | 2.6 % |
| Other | 5.2 % |

We asked those respondents who did not use GM to pick all of the reasons that they did not use it (table 6).

**Table 6. Reasons Google Moderator was not used**

| I prefer to ask my questions out loud | 11.2 % |
|---|---|
| I did not have a Google account | 7.1 % |
| I did not bring my laptop | 18.4 % |
| I did not understand how to use Google Moderator | 3.1 % |
| I did not have a question | 28.6 % |
| I was preoccupied with the lecture | 21.4 % |
| I did not attend class | 7.1 % |
| Other | 3.1 % |

#### 4.2.2 Observations

We observed the number of students with a laptop was roughly one third of the class, and there was not a significant increase during the study. This did not exclude the students without laptops from the study, as most students carry smart phones; however, GM's smartphone interface is not very user friendly and this may have impacted the number of students who used it.

We observed the question-ranking feature of GM was not providing enough impact, as there were not enough students in the class using it. As a result of this we encountered persistent questions sitting on top of the teacher's question queue, preventing the teacher from noticing if any new questions were submitted. A related issue we observed was GM has no explicit notification abilities, so when a new question was a submitted, the teacher was not explicitly made aware. For some questions this created a lack of context. For example, one student submitted a question "why" in response to something the teacher said. This question at the time of submission was a valid question, however, the teacher did not see it for roughly twenty minutes, which meant all context was lost.

In class *C* the instructor asked students to submit questions to GM that could be answered at the start of the next lecture, and the instructor set aside classroom time for the students to do this. This created a large influx of questions at the end of class. We observed that any time an instructor actively encouraged the use of GM by allocating classroom time for its use, the number of questions submitted increased.

### 4.2.3 Student Comments
Many students reported liking the tool. One student mentioned, "It seems very helpful for people who have questions but are afraid to speak up in class." Another student mentioned, "I like this option, especially for large lecture classes where a question can prove stressful." Some students expressed frustration: one student commented, "a lot of questions went unanswered." Another student commented that they would like to be able to submit questions by texting.

### 4.2.4 Instructor Feedback
Instructor *DE*, the one conducting the flipped classroom, initially wanted to use GM on their cellphone. The instructor had difficulty doing this, finding the phone interface to be too cumbersome and eventually stopped using it. Instructor *DE* also commented during the study that they felt GM would not be effective due to the number of individual questions they already receive.

Instructor *C* was very impressed with GM and was very enthusiastic about its use. However, the instructor GM was being used in their other class, as they felt the level of professionalism among the other students was higher. They also felt that their other class would be more mature and able to ask questions. Instructor *C* also mentioned that the timing of the study might have been an issue for her class as the material being studied at that time did not generate questions easily. Instructor *C* mentioned that they would like to use GM again in the future.

Instructor *AB* was overall happy with GM. They mentioned that they liked how students were submitting questions overnight to be answered the next day in class.

## 5. DISCUSSION
Our original model of an LAQQ system supported six features: (1) live feedback, (2) anonymity, (3) a question queue, (4) question-ranking, (5) notifications, and (6) accessibility. Through this study our understanding of an LAQQ system has broadened to include additional support for (7) shared memory, (8) context, and (9) class structure. We discuss this broader set below, in relation to existing literature on technology use in the classroom.

**Live Feedback**
Our study found that support for live feedback can provide instructors with immediate feedback on the comprehension level of the students. When instructors were looking at the external displays, they were able to immediately address the student's questions. These results are in line with previous studies providing external display [2,7,9]. These findings confirm that live feedback is essential for an LAQQ system.

**Anonymity**
Our findings suggest that providing an anonymous environment for students affords them comfort when addressing their lecturer in the classroom. Similar to the other tools in the literature [2,6,7,9] GM had anonymity built in. Whereas the previous studies assumed anonymity is desired in their systems, our study specifically investigated the role that anonymity played in the classroom environment. We found that 41% of our survey respondents were uncomfortable asking questions, and 93% of them said that they would be more comfortable if they could ask questions anonymously. This confirms Tammy et al.'s finding that anonymity is a solution to student apprehension [9]. Table 3 suggests that the main reasons for apprehension are a fear of being judged by others for asking a dumb question, and to a lesser extent, a difficulty wording questions. These findings confirm that an LAQQ system must support anonymity.

**Question Queue**
Our study showed a question queue provides an effective information feed of student concerns to be addressed in turn by the instructor. During the course of our study, the instructors referred to the contents of the question queues, to gather information about their student's level of understanding from previous classes. The importance of a question queue as a feature of an LAQQ is supported by the implementation of questions queues in the related works [2,6,7,9].

**Question Ranking**
Question ranking can provide an LAQQ system self-moderation and student-driven question-popularity ranking if adopted by the classroom by enough students. However, the question-ranking feature in GM was not effective during our study. Not enough students were using GM, and an even smaller number of students were ranking questions. With no one using the feature it did not provide an accurate measure on question popularity and it did not serve to regulate inappropriate or irrelevant submissions.

Two of the instructors in our study wanted to use question ranking to determine what questions to answer. Bergstrom et al. found that as their system did not provide question ranking, students were bumping relevant questions out of the question queue by submitting questions such as: "+1" and "me too" [2]. So while question ranking was not successful for GM in our, we still suggest it as a necessary feature for an LAQQ.

**Notifications**
The awareness model currently used by lecturers needs to change to accommodate notifications other than just raised hands, and an LAQQ system needs to support notifications to fit this new model. We found that some of the questions posted during our study were not noticed by the lecturers due to too persistent questions and a lack of notification to provide awareness when questions initially come in. This was noted by one of the survey respondents, "Lots of questions went unanswered." Question persistence is useful for shared memory, but only when the persistence does not hinder awareness. An LAQQ system must support awareness through notification for both the lecturer and students.

**Accessibility**
The level of accessibility students and lecturers have to an LAQQ is important for its success in the classroom. Although only one third of the students in our study had laptops, most had smartphones. However, the mobile interface for GM was not easy to use. The instructor who initially began using GM on their smartphone eventually stopped using it due to usability issues. Ratto et al. had similar issues with the software on the PDAs they handed out [7]. This study has shown that accessibility of an LAQQ is important for its success. In particular, an LAQQ should be usable from any popular platform.

**Shared Memory**

An LAQQ system can provide additional classroom support – aside from solving apprehension – by archiving student questions in a database that would act as shared memory for the classroom. Nunamaker et al. propose that the "failure to remember the contributions of others" is a source of group process loss [5]. When students ask questions in class they are contributing to the class by slowing the instructor down, highlighting confusing areas, and addressing questions that other students may have been having difficulty phrasing. If these questions could be stored in an LAQQ they could continue to provide value to the class and the LAQQ could act as a shared class memory. It is for these reasons that we recommend an LAQQ support group shared memory by supporting categorization and storage of questions.

**Context**

Support for context-rich questions must be provided by an LAQQ system to mirror or surpass the level of context provided by the conventional method of asking questions out loud. Tammy et al. mentions shared context between lecturer and students as being a feature that a computer-mediated feedback system should support [9]. We initially thought that the timing of the question submission along with the text format of the question would provide enough context support in GM. This did not turn out to be the case largely due to the lack of notification in GM.

If a lecturer did not see a question come in right away the timing of that question was lost, and if the question does not provide suitable context in its phrasing then it becomes difficult to answer. For example, the question "why" was submitted during one of the classes. At the time of submission this question was valid and related to course material, however, when the lecturer saw the question several minutes after submission they were unable to answer the question as it had lost all context. Although notification support in an LAQQ will help to provide context, we recommend investigating additional methods.

**Class Structure**

Integrating an LAQQ into the classroom setting requires the classroom environment to change. This is consistent with Bannon and Schmidt's proposal that "changes in technology induce changes in the work organization" [6]. Instructor *AB* allotted 5 minutes at the beginning of their class to answer questions from GM. Instructor *C* allotted 10 minutes at the end of their lecture to allow for posting and answering of GM questions. Instructor *DE* conducted a flipped class, which also allowed for students to post to GM. These are examples of the technical system inducing changes in the classroom environment.

Our survey showed that a quarter of the students did not ask a question on GM because they were preoccupied with the lecture. Additionally, it was during the period that instructor *C* provided time for their students to submit questions at the end of class that the most questions for that class were submitted. This suggests to us that the class structure can have an effect on the LAQQ system, and if the class does not provide enough time for students to use the system then they might not use it.

Nunamaker et al. propose attention blocking and information overload as sources of group process losses [5]. Unfortunately an LAQQ does not solve these problems directly; however, if the class structure changes to allow for time to submit questions to the LAQQ it will help to reduce attention blocking. If an LAQQ provides adequate support for shared memory and context it can serve as a resource for students to look back on thereby helping to minimize the effects of information overload.

## 6. CONCLUSION

Overall, our findings indicate that student apprehension is a significant problem in large classrooms and that anonymity is a desired solution. Providing a system for students to anonymously submit questions to their instructor will allow students that are otherwise too afraid to speak in front of their peers to receive answers on their questions. Our findings also indicate that an LAQQ should provide support for notification and context. Additionally an LAQQ should be accessible to all students on multiple platforms. Finally, in order for an LAQQ system to be successful in a classroom it is necessary for the class organizational structure to change and for the instructors to fully adopt the LAQQ system and encourage its use.

## 7. ACKNOWLEDGMENTS

We'd like to thank our teammates: Warren Spencer, Chris Cook, and Ian Brown for their help conducting and planning the study. We'd also like to thank the participating instructors for allowing us to conduct our study in their classrooms.

## 8. REFRENCES